# Dynamically tunable moiré Rydberg excitons in a monolayer semiconductor on twisted bilayer graphene

Minhao He[1†], Jiaqi Cai[1†], Huiyuan Zheng[2†], Eric Seewald[3], Takashi Taniguchi[4], Kenji Watanabe[5], Jiaqiang Yan[6,7], Matthew Yankowitz[1,8], Abhay Pasupathy[3], Wang Yao[2,9#], Xiaodong Xu[1,8#]

[1]Department of Physics, University of Washington, Seattle, Washington, 98195, USA
[2]Department of Physics, University of Hong Kong, Hong Kong, China
[3]Department of Physics, Columbia University, New York, NY, 10027, USA
[4]International Center for Materials Nanoarchitectonics, National Institute for Materials Science, 1-1 Namiki, Tsukuba 305-0044, Japan
[5]Research Center for Functional Materials, National Institute for Materials Science, 1-1 Namiki, Tsukuba 305-0044, Japan
[6]Materials Science and Technology Division, Oak Ridge National Laboratory, Oak Ridge, Tennessee, 37831, USA
[7]Department of Materials Science and Engineering, University of Tennessee, Knoxville, Tennessee, 37996, USA
[8]Department of Materials Science and Engineering, University of Washington, Seattle, Washington, 98195, USA
[9]HKU-UCAS Joint Institute of Theoretical and Computational Physics at Hong Kong, China

[†] These authors contributed equally to the work.
[#]Correspondence to: wangyao@hku.hk; xuxd@uw.edu

**Abstract: Moiré excitons are emergent optical excitations in 2D semiconductors with deep moiré superlattice potentials. While these excitations have been realized in several platforms, a system with dynamically tunable moiré potential to tailor the moiré exciton properties is yet to be realized. Here, we present a continuously tunable moiré potential in a monolayer $WSe_2$ that is enabled by its proximity to twisted bilayer graphene (TBG) near the magic-angle. Due to its flat electronic bands, charge distribution is highly localized and forms a triangular lattice in TBG. Tuning the local charge density via electrostatic gating, TBG thus provides a spatially varying and dynamically tunable dielectric superlattice for modulating monolayer exciton wavefunctions. By performing optical reflection spectroscopy, we observe emergent moiré exciton Rydberg branches in monolayer $WSe_2$ with increased energy splitting upon doping TBG. The twist-angle dependence reveals that the observation is due to a hybridization between bright and dark Rydberg states enabled by the moiré potential. Further, at the magic-angle near ≈1.1°, the moiré Rydberg excitons form a sawtooth pattern with doping owing to the formation of strongly correlated states in the TBG. Our study provides a new platform for engineering moiré excitons as well as optical accessibility to the electronic states with small correlation gaps in TBG.**

**Maintext**

The ability to engineer and control the wave function of quasi-particles is of critical importance for quantum science and technologies. Recently, the creation of synthetic moiré quantum matter formed by stacking layered van der Waals materials provides a new and powerful means for achieving such control[1–3]. An example are the moiré excitons found in transition metal dichalcogenide (TMD) moiré superlattices[4–8], which act as novel optical excitations residing in a smoothly varying periodical moiré potential. Compared to free 2D excitons, the introduction of the moiré potential can drastically modify the properties, with effects including new optical resonances due to zone folding[6,9], optical selection rules determined by local sublattice symmetry[5,7], and quantum optical effects from deep potential confinement[10,11]. In addition to single trap physics, the periodic superlattice potential enables collective effects such as topological excitons[12], excitonic insulators[13], exciton-correlated electron bound state[14], and dipolar exciton solids[15,16].

Since the moiré potential and its lattice geometry play a critical role in the physical properties of moiré excitons, it will be desirable to achieve dynamic tunability and even an on/off switch with which to control the moiré potential. While the spatial periodicity can be engineered by lattice mismatch and/or twist angle during sample fabrication, achieving wide-range continuous tunability of the moiré potential depth remains challenging owing to the intertwined electronic and structural origin of the moiré potential[17,18]. An alternative solution is to directly engineer the dielectric environment of monolayer TMDs. The Wannier-Mott nature of excitons in monolayer TMDs makes them extremely sensitive to their dielectric environment[19,20], especially so for the Rydberg excitons with large Bohr radius[19,21,22]. This effect has been successfully utilized to engineer exciton binding energy by substrate design[20,23]. As a result, the excitonic moiré potential, arising from the periodic modulation of both the quasiparticle band gap and exciton binding energy, can be engineered in monolayer TMDs placed directly onto a dielectric superlattice. Recent work has demonstrated the periodic dielectric screening effect in a heterostructure of monolayer $WSe_2$ and a moiré superlattice of graphene on $hBN$[24], where the presence of moiré potential is revealed by high energy replica of the 2s exciton. Moreover, periodic dielectric screening of electron-hole Coulomb exchange can also be explored for engineering spatial valley textures to enable novel excitonic functionalities[25].

Here, we employ a twisted bilayer graphene (TBG) as a functional substrate to engineer a dynamically tunable moiré potential in monolayer TMDs. We show that this platform enables the control of excitonic orbitals, and thus a tunable hybridization between bright and dark Rydberg excitons, manifested as brightened exciton states versus doping. Figure 1a depicts the device structure. For TBG near magic-angle, the flat electronic bands enable the generation of a spatially modulated dielectric constant, with its strength proportional to the localized but tunable charge density. Taking a TBG sample with a twist angle of 0.79° as an example, scanning tunneling microscopy (STM) imaging shows the local atomic registry varying from AA sites (carbon atoms in the two layers overlap vertically) to AB/BA sites (Bernal stacking) within single moiré unit cell[26–28] (Figure 1b). The charge distribution is highly localized on the AA sites and forms a triangular lattice, as further seen in a map of the local density of states (LDOS) (Figure 1c). Gating the TBG enables direct control of the charge density at the AA sites, while always maintaining the same overall spatial pattern. Interfacing TBG with monolayer TMD could thus realize a dynamically tunable and spatially periodic dielectric environment for the monolayer excitons.

**Observation of moiré Rydberg excitons**
We have studied a total of four WSe$_2$/TBG devices here, in which the twist angle of TBG ranges from close to 0° (i.e., nearly complete relaxation to Bernal stacking) up to the magic angle (≈1.1°) (Extended Data Figure 1). The devices are constructed in dual-gated Hall bar geometry, allowing for individual control over displacement field $D$ and carrier density $n$ (Figure 1a, See Method for details). Electrical transport measurements are first performed to characterize each device and determine the twist angle $\theta$ of TBG (Extended Data Figures 2-4). We then employ optical reflection spectroscopy to study the excitons in the monolayer WSe$_2$. All optical spectroscopy data are acquired at a nominal temperature of 5 K, unless otherwise specified.

In Figure 2a, we first present the gate-dependent optical reflectance contrast ($\Delta R/R$) of a reference sample, where the monolayer WSe$_2$ is adjacent to a TBG with a nearly relaxed twist angle (i.e., no moiré effects, see Extended Data Figure 2 for transport measurements). Compared to the reflectance contrast of isolated monolayer WSe$_2$ (Extended Data Figure 5)[29–31], gate-dependent $\Delta R/R$ of the WSe$_2$/TBG heterostructure is dominated by the 1s exciton at 1.72 eV with negligible energy shift as a function of the doping. This is consistent with the band alignment of graphene and WSe$_2$: the Fermi energy can be gate-tuned within the TBG flat band, but always remains deep inside the WSe$_2$ band gap[24] (Figure 2b inset).

Moreover, a 2s exciton is observed at ~80 meV above the 1s exciton at zero doping. The 1s - 2s exciton splitting, which implicitly connects to the exciton binding energy, is much smaller than the 130 meV splitting observed in monolayer WSe$_2$ encapsulated by hBN (Extended Data Figure 5). The reduction of the exciton binding energy and suppression of the oscillator strength of the 2s exciton[20] points to a strong dielectric screening effect from the TBG[20]. The dielectric screening effect is enhanced as charge carriers are electrostatically doped into the TBG, as indicated by the continuous suppression of the oscillator strength of the 2s exciton and the redshift of its peak energy.

We next show optical reflection measurements of a monolayer WSe$_2$ on TBG with a twist angle $\theta$ = 0.8° (Fig. 2c), with $\theta$ determined from the Hall resistance measurements (Fig. 2d). The tunable moiré dielectric screening effect is revealed by the emergence of exciton Rydberg states upon gating. As seen in Fig. 2c, the 2s exciton peak is the only Rydberg exciton observed at charge neutrality, consistent with the reference sample. A strong excitonic peak emerges at slightly higher energy upon doping the TBG to $n \approx \pm 0.5 \times 10^{12}$ cm$^{-2}$, and disperses to lower energy as the TBG is further doped. This is followed by the emergence of a second, weaker excitonic peak at $n \approx 1.5 \times 10^{12}$ cm$^{-2}$ on the electron-doped side that disperses only weakly with TBG doping (Extended Data Figure 6). The energy splitting between the Rydberg excitons also increases with doping, highlighted in the spectral linecuts taken at selected dopings (Fig. 2e). These emergent peaks are reproducible across the whole sample and are robust up to ~150 K (Figure 2f). Similar phenomena are also observed in another device with TBG $\theta$ = 0.93° (Extended Data Figure 7).

Figure 3a shows the extracted relative energy $\Delta E$ of the Rydberg states with respect to their corresponding 1s states, measured in WSe$_2$/TBG devices with different twist angles. At zero doping, a single 2s state is observed 80 meV above the 1s state, independent of the TBG twist angle. This reveals the weak moiré modulation of the dielectric screening when the TBG is held at charge neutrality, consistent with the expectation of a uniform underlying charge distribution in neutral TBG. In contrast, devices with a well-defined TBG moiré superlattice rapidly develop a strong moiré potential with doping as a result of the periodically modulated charge distribution in TBG (Fig. 1c). Whereas small-angle TBG with nearly-Bernal stacking exhibits only a 2s exciton, our observation of multiple gate-tunable excitons in the samples with larger twist angle is consistent with the formation of moiré Rydberg exciton states. This clearly demonstrates the importance of the underlying periodic dielectric screening effect from the TBG moiré potential.

**Hybridization of excitons with distinct orbitals**

We now turn to determining the origin of these new Rydberg states (see details in Extended Data Note). Excitonic zone-folding effects due to the moiré superlattice provide one possible explanation, as explored in previous works[6,9,24]. In brief, the periodic potential folds the original exciton dispersion into a moiré mini-Brillion zone. When the folded bands are within the light cone with finite optical oscillator strength, new excitons in the form of Umklapp states arise at higher energy. To examine this possible explanation, we extract the energy difference between two lowest Rydberg states. Figure 3b shows that the energy difference from various twist angles nearly collapse onto a single curve as function of doping. Since the energy separation of an Umklapp state with respect to the ground state is proportional to $\theta^2$, our observation of an energy separation that is nearly independent of $\theta$ does not support such a possibility[24].

In fact, the moiré potential effect in this zone-folding picture only affects the exciton center-of-mass motion, assuming that the Rydberg orbitals of relative electron-hole motion remain intact. This assumption breaks down once the Bohr radius of Rydberg states[32,33] becomes comparable to the moiré periodicity. The rapidly varying moiré potential introduces off-diagonal coupling between the otherwise orthogonal Rydberg states. Therefore, the hybridization between excitonic Rydberg states that are close in energy becomes appreciable. Moreover, through such hybridization, dark Rydberg states can acquire finite oscillator strength from a hybridized component of bright ones. As a result, the exciton hybridization effect driven by the enhanced moiré potential of the doped TBG can generate new, bright excitonic features in the optical spectrum as the corresponding dark states acquire oscillator strength.

We now present a phenomenological model to describe the exciton hybridization effect with doping (see Methods). We consider that a moiré potential couples the 2s state to a nearby optically dark or dim state (i.e., 3s and 4s Rydberg states[32]) with a coupling strength $t$ that grows with doping (c.f. the schematic Fig. 3c). Fig. 3d shows a calculation of the spectrum of the excitonic states (oscillator strength) as function of doping of the TBG, reproducing the main features of our experiment. The calculation shows that moiré-hybridized excitonic states brighten as the moiré potential is enhanced. Moreover, the splitting of these hybridized states with the 2s exciton also increases, reflecting the increased coupling strength.

**Optical sensing of strongly correlated states in magic-angle TBG**
The highly localized charge density in TBG is fundamental in realizing the gate-tunable periodic potential in its adjacent monolayer semiconductor. While this statement holds true for a large range of TBG twist angles, a narrower range near the magic-angle of 1.1° is of particular interest, at which the interlayer hopping interference creates extremely flat low-energy electronic bands[34]. The kinetic energy of the electrons is suppressed within these flat bands, and the system becomes dominated by strong Coulomb interactions. A rich phase diagram of strongly correlated states have been reported in magic-angle TBG [35–40], including correlated insulators, superconductors, orbital magnetism and the quantum anomalous Hall effect.

Figure 4a shows the $\Delta R/R$ of moiré Rydberg excitons of a monolayer $WSe_2$ adjacent to magic-angle TBG with $\theta = 1.03°$, measured at $T = 2$ K. Unlike the continuous energy redshift as discussed in Fig. 2, the 2s exciton follows a sawtooth pattern in which its energy temporarily blueshifts near certain charge densities but exhibits an overall redshift trend as the TBG is doped (see also Extended Data Figure 8). The blueshifts of the 2s exciton are accompanied by an enhancement in its oscillator strength enhancement, which combined indicates a reduced dielectric screening at these specific carrier densities. To help understand this exciton spectrum, we conduct electrical transport measurements in the same device as shown in the top panel of Figure 4b (further transport characterization is shown in Extended Data Figure 4). We highlight the high quality of our sample with the superconductivity state emergent below $T = 1$ K at $\nu = -(2+\delta)$, as seen in the inset. Within the flat bands between $\nu = \pm 4$, the $WSe_2$/TBG is characterized by a cascade of strongly correlated states, either insulating or weakly resistive, at almost all integer filling factor $\nu = -3, -2, 0, 1, 2, 3$, consistent with previous reports[41–43]. A variety of symmetry-broken states have been proposed as ground state candidates in magic-angle TBG[44], which is further complicated by induced spin-orbit coupling from the proximate $WSe_2$[41–43] in our device. The exact nature of these correlated states remains an open question with active investigation.

We observe a clear connection between the sample resistance $R_{xx}$ and the 2s-1s splitting $\Delta E$ (Figure 4b). We identify a one-to-one correspondence of the maxima of energy blueshifts and oscillator strength of the 2s exciton to the correlated insulators at $\nu = +1, +2,$ and $+3$ seen in transport. The correlated insulating state in the hole-doped states are not as clear as the electron doped side. In the static limit, Thomas-Fermi screening establishes a direct relationship between dielectric constant ε and electronic compressibility $\partial n/\partial \mu$. As a result, the peak of 2s exciton spectrum can serve as an optical sensor of local compressibility. We note that further theoretical work is needed to establish a quantitative relation bridging the exciton spectrum to the dielectric superlattice, as well as the electronic compressibility.

**Conclusions**
In conclusion, our work develops a continuously tunable moiré potential for monolayer semiconductors, arising due to pronounced dielectric screening from the unique charge distribution in TBG. This creates opportunities to engineer trapping arrays for monolayer excitons with selectivity to their specific Rydberg orbitals, and to tailor exciton dynamics and many-body interactions using this periodic potential landscape. Rydberg excitons can also be exploited as a dynamical sensor to enrich the toolbox for probing equilibrium states and dynamics of correlated phenomena of interest in magic-angle TBG, such as charge order, magnetic order and superconductivity.

**Methods:**

**Sample fabrication**

Heterostructures of graphite/hBN/WSe$_2$/TBG/hBN/graphite are assembled using a standard dry-transfer technique with a PC/PDMS (polycarbonate/polydimethylsiloxane) stamp and transferred onto a Si/SiO$_2$ wafer. TBG stack is fabricated by using the tear-and-stack method. We use 3-5 nm graphite as bottom gates whereas 3-5 layers graphene are used for top gate to minimize its optical absorption. We have no intentional control over the twist angle between WSe2 and its adjacent graphene layer during the stacking process. CHF$_3$/O$_2$ and O$_2$ plasma etching followed by electron beam lithography are used to define a Hall bar geometry, and Cr/Au contacts are finally added using electron beam evaporation.

**Transport measurements**

Transport measurements were conducted in a Quantum Design PPMS system. Measurements are performed with a 1-5 nA a.c. excitation current at either 13.3 Hz or 13.7 Hz. The current and voltage are pre-amplified by DL 1211 and SR560 respectively, and then read out by SR830/SR860 lock-in amplifiers. Gate voltages are supplied by either NI DAQ or Keithley 2450. A global Si gate is sometimes used to reduce contact resistance issue.

The dual-gate device geometry enables independent control of $D$ and $n$. The relationship between the top and bottom gate voltages, $V_t$, $V_b$, and n and D are given by $n = (V_t C_t + V_b C_b)/e$ and $D = (V_t C_t - V_b C_b)/2\epsilon_0$, where $C_t$ and $C_b$ are the top and bottom gate capacitances per unit area and $\epsilon_0$ is the vacuum permittivity.

The filling factor $\nu$ is defined as the number of electrons per moiré unit cell. Full filling of the eight-fold degenerate flat bands in TBG corresponds to 4 electrons (holes) per moiré unit cell, $\nu$ = +4 (-4). The twist angle was first determined by measuring the carrier density corresponding to the band insulators at $\nu = \pm 4$, following the relationship $n = 8\theta^2/\sqrt{3}a^2$, where $a$ = 0.246 nm is the graphene lattice constant. For the TBG at magic angle, it is further confirmed by fitting the observed quantum Hall states and Chern insulators with the allowed Hofstadter states in the Wannier diagram.

**Reflectance contrast spectroscopy**

Optical reflection measurements are performed using a home-built confocal microscope. The sample was mounted either on the cold head of a close-cycled Montana cryostat with temperature kept at 5 K, or inside a He-exchange-gas cooled cryostat (Quantum Design OptiCool) with temperature kept at 2 K. A halogen lamp (Thorlabs SLS201L) is used as a white light source. The output of the white light is coupled to a single-mode fiber and then collimated with a triplet collimator. The beam was then focused to a beam waist of ~1 μm onto the sample by a 40x objective lens (NA = 0.6). The excitation power of the white light is kept below 10 nW. Reflection signals are collected and directed into a spectrometer, where they are dispersed by a diffraction grating and detected on a silicon charge-coupled-device (Princeton PyLoN).

Reflectance contrast $\Delta R/R$ is defined as $\Delta R/R = (R - R_0)/R_0$, where $R$ denotes the reflection spectrum on the sample and $R_0$ denotes the background reflection spectrum. For $\Delta R/R$ measurements on monolayer WSe$_2$ (Extended Data Figure 5), the background reflection spectrum is taken at an adjacent spot without WSe$_2$. For $\Delta R/R$ measurements on the WSe$_2$/TBG heterostructure, the signal from the Rydberg states is much weaker. Hence the background reflection spectrum for those measurements is taken at the same spot on the sample, but with sample gate tuned to a high doping level. To minimize the overtime drift of the $\Delta R/R$ background, $R_0$ spectrum is taken periodically[45] (approximately every 10 mins).

**Scanning Tunneling Microscopy (STM)**
TBG samples are prepared using the tear-and-stack method using PPC as a polymer to sequentially pick up hexagonal boron nitride (hBN), half of a piece of graphene followed by the second half with a twist angle. This structure is flipped over and placed on an Si/SiO$_2$ chip without any further processing of the polymer. Direct contact is made to the TBG via micro-soldering with Field's metal at a temperature of 65 C in ambient conditions. No further annealing or processing of samples is performed prior to measurement. UHV STM and spectroscopy measurements are carried out in a home-built UHV STM at a temperature of 5 K. All tips used in this study are prepared on Au (111) and calibrated spectroscopically prior to measurements on graphene. Several freshly prepared tips were used in this study to ensure consistency and accuracy of the findings.

**Exciton calculation**
A phenomenological calculation of the Rydberg exciton hybridization is provided here. We assume the $Ns$ Rydberg states $\psi^{(Ns)}$ have the energy $E^{(Ns)}(n)$ (Extended Data Figure 10), dependent on the average carrier density n in the TBG layers. The exciton binding is actually sensitive to the local charge distribution. With the sharp variation of the charge density distribution in TBG in length scale already comparable to exciton radius in the monolayer WSe$_2$, the Rydberg states in the pristine monolayer are no longer eigenstates of the Coulomb binding in the presence of doped TBG. Hybridization of the Rydberg states that are close by in energy becomes significant. We consider the coupling of the 2s state with two nearby states (3s, 4s), with matrix elements $t_{23}(n)$ and $t_{34}(n)$ dependent on the average carrier density. The energy $E_N$ of the hybridized state can be calculated by solving:

$$H = \begin{bmatrix} E^{(2s)} & t_{23} & 0 \\ t_{23}^* & E^{(3s)} & t_{34} \\ 0 & t_{34}^* & E^{(4s)} \end{bmatrix}$$

The spectral function of the hybridized state can be calculated:

$$A(E,n) = -\frac{1}{\pi} Im \sum_N \frac{|c_N|^2}{E - E_N + i\gamma}$$

The simulated spectrum show in Figure 3d is obtained using the carrier density dependence of the parameters given in Extended Data Figure 10. Further discussion on the moiré potential induced exciton hybridization is detailed in Extended Data Note.

**Acknowledgments:** We thank Jia Leo Li for helpful discussions. The work is mainly supported by DoE BES under award DE-SC0018171. Sample fabrication is partially supported by ARO MURI program (grant no. W911NF-18-1-0431). Electrical transport measurement is partially supported by the U.S. National Science Foundation through the UW Molecular Engineering Materials Center (MEM·C), a Materials Research Science and Engineering Center (DMR-1719797). Scanning tunneling microscopy/spectroscopy measurement is supported by the Center on Programmable Quantum Materials, an Energy Frontier Research Center funded by the U.S. Department of Energy (DOE), Office of Science, Basic Energy Sciences (BES), under award DE-

SC0019443. Work at HKU is supported by the Research Grants Council of Hong Kong SAR (AoE/P-701/20, HKU SRFS2122-7S05). WY acknowledges support from Tencent Foundation. K.W. and T.T. acknowledge support from the JSPS KAKENHI (Grant Numbers 19H05790, 20H00354 and 21H05233). XX acknowledges support from the State of Washington funded Clean Energy Institute and from the Boeing Distinguished Professorship in Physics.

**Author Contributions:** M.H., J.C. performed the transport and optical reflection measurements, under the supervision of X.X, M.Y. M.H. fabricated the samples. M.H., J.C., M.Y., W.Y., and X.X., analyzed and interpreted the results. H.Z., W.Y. performed theoretical calculations. E.S., A.P. performed STM measurements and analyzed the results. J.Y. synthesized and characterized the bulk WSe2 crystals. T.T. and K.W. synthesized the hBN crystals. M.H., X.X., W.Y., J.C. and H.Z. wrote the paper with input from all authors. All authors discussed the results.

**Competing Interests:** The authors declare no competing interests.

**Data Availability**: All data that support the plots within this paper and other findings of this study are available from the corresponding author upon reasonable request. Source data are provided with this paper.

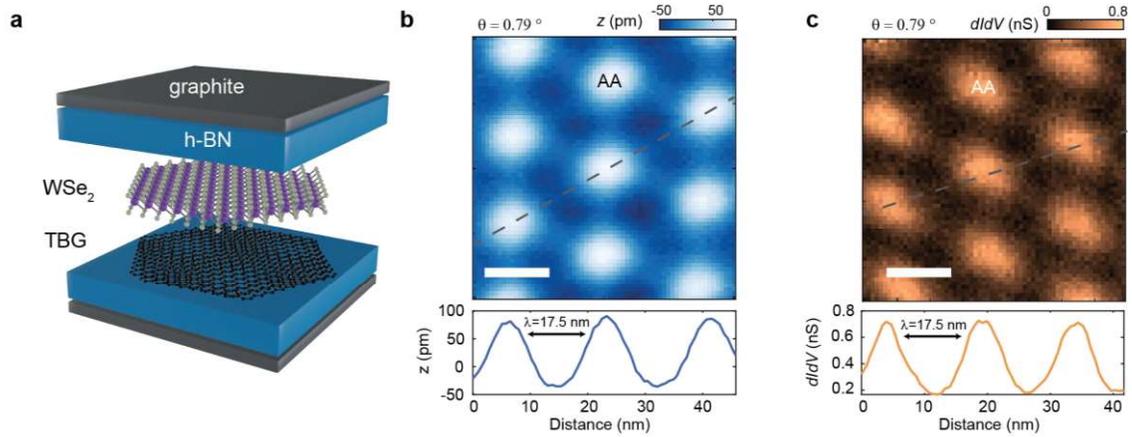

**Figure 1. Localized charge distribution in twisted bilayer graphene and its periodic screening of an adjacent WSe₂. a,** Schematic of the device structure, where a monolayer WSe$_2$ is stacked on top of the TBG. The heterostructure then is further encapsulated with h-BN on both side with top and bottom graphite gates. **b,** Scanning tunneling microscopy topography image of a $\theta = 0.79°$ twisted bilayer graphene (TBG) sample. The scale bar is 10 nm. The bottom panel shows the linecut of the topography image along the dashed line. **c,** Local density of states (LDOS) at charge neutrality in the valence flat band on the same TBG sample. Spectroscopic measurements are performed with a junction normalization of V=0.5 V, I=100 pA. Under these conditions, the STM tip follows the topographic height of the sample well and the contrast arises from spectroscopic variations.

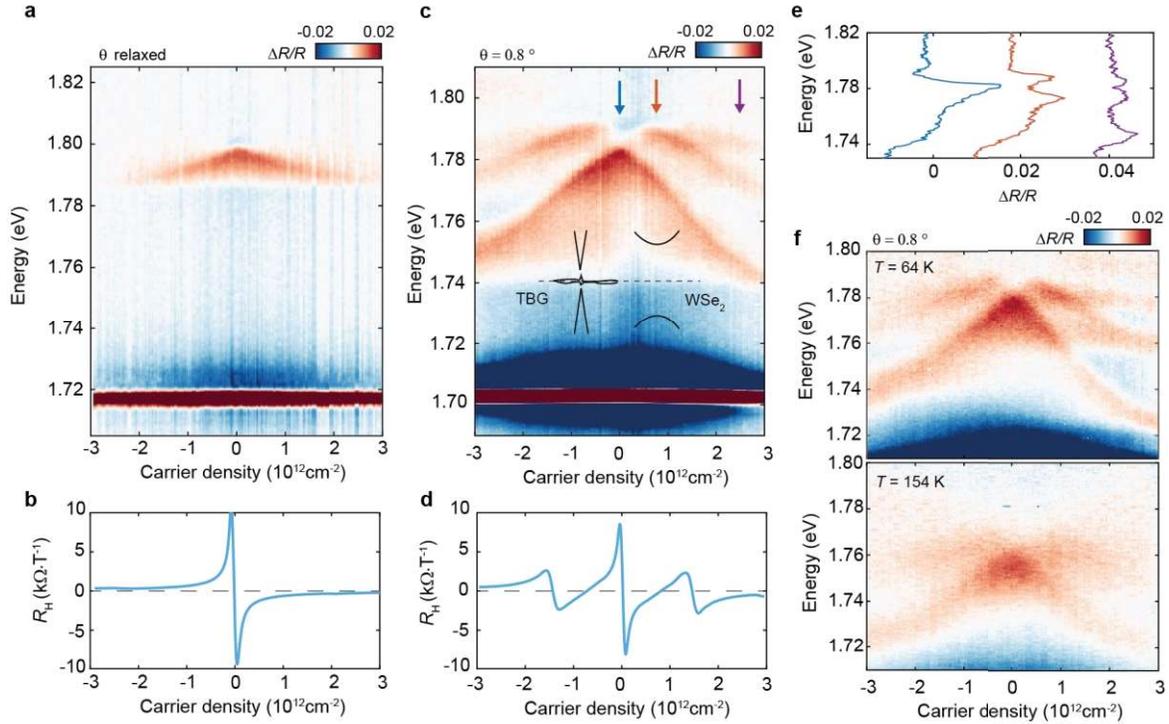

**Figure 2. Moiré Rydberg excitons in monolayer WSe$_2$ on twisted bilayer graphene. a,c,** Reflectance contrast $\Delta R/R$ as a function of carrier density of a monolayer WSe$_2$ on TBG with **a,** twist angle $\theta$ relaxed; **c,** twist angle $\theta = 0.8°$. The band alignment of the monolayer WSe$_2$ and TBG is shown in **c** inset. Data is taken at 5K. **b,d,** Hall coefficient (R$_H$) of the TBG in the device shown in panels **a** and **c** correspondingly. **e,** Linecuts of $\Delta R/R$ in the device with TBG $\theta = 0.8°$, at carrier density labeled by arrows in panel **c**, which are n = 0, 0.73 x 10$^{12}$, 2.43 x 10$^{12}$ cm$^{-2}$ for the blue, orange and purple lines respectively. **f,** $\Delta R/R$ of the same device with TBG $\theta = 0.8°$ measured at temperature $T = 64$ K and $T = 154$ K.

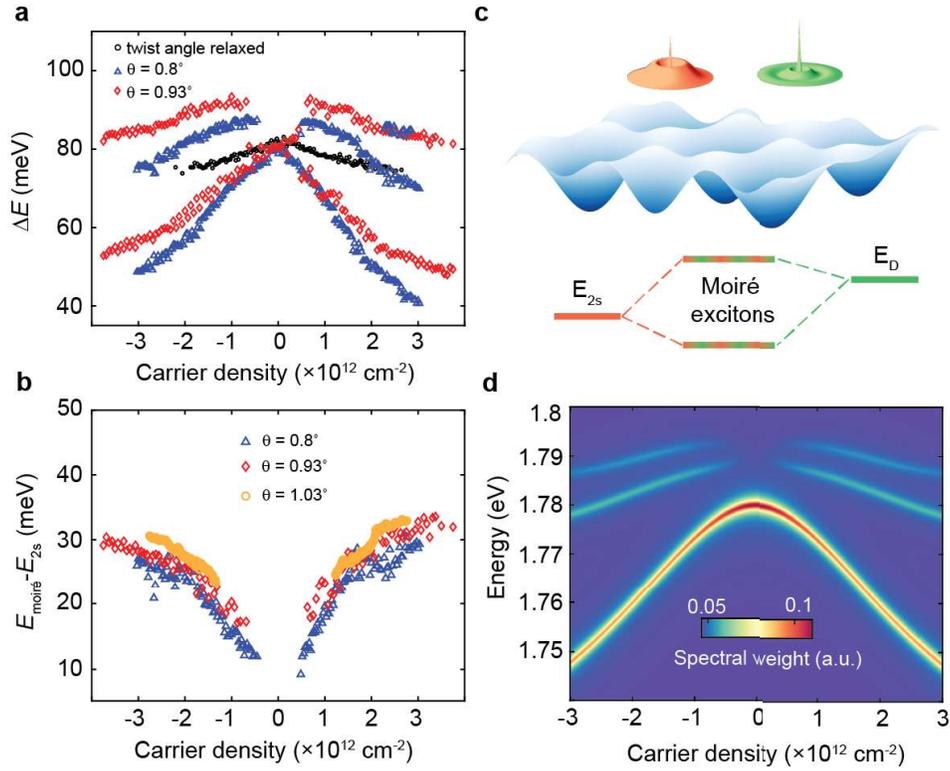

**Figure 3. Tunable Rydberg exciton hybridization. a,** Relative energy $\Delta E$ of the Rydberg excitons with respect to 1S exciton in WSe$_2$/TBG at different twist angles $\theta$, i.e. $\Delta E = E - E_{1s}$. Here E and E$_{1s}$ are the Rydberg excitons and 1S exciton energy, respectively. **b,** Energy difference between the higher energy moiré excitons and the 2s exciton, plotted as a function of TBG doping in samples with different twist angle $\theta$. **c,** Schematic of the 2s exciton and another dark or dim exciton (3s Rydberg exciton in the model) sitting in a rapidly varying moiré potential, with periodicity comparable with the Bohr radius of the excitons themselves. The bottom panel shows the exciton hybridization effect that results in moiré excitons inheriting the spectral weight of the original 2s exciton state. **d,** Simulated spectral weight of the hybridized Rydberg excitons brightened by the moiré potential. The calculation details can be found in Extended Data Note.

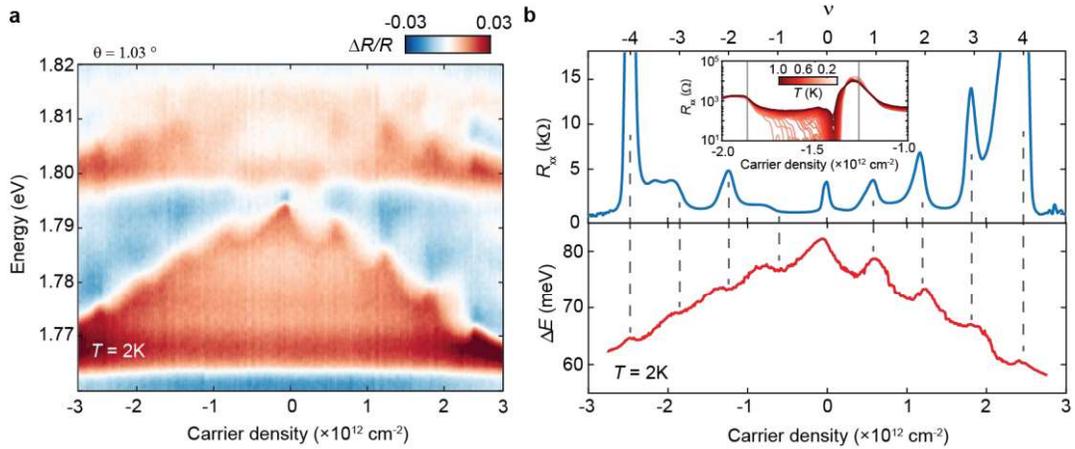

**Figure 4. Rydberg exciton sensing of strongly correlated states in magic-angle TBG. a,** Reflectance contrast $\Delta R/R$ of a monolayer WSe$_2$ sitting on top of a magic-angle twisted bilayer graphene with $\theta = 1.03°$. Data is taken at $T = 2$ K. **b,** Resistance ($R_{xx}$) of the TBG at $T = 2$ K (top) and the relative energy $\Delta E$ of the 2s exciton ($\Delta E = E_{2s} - E_{1s}$) extracted from panel **a** (bottom), as function of carrier density. The inset shows the temperature dependence of the TBG $R_{xx}$ in the vicinity of filling factor $\nu = -2$ with emergent superconductivity at temperature below $T = 1$ K, demonstrating the high quality of the sample.

# Extended Data for:

# Dynamically tunable moiré Rydberg excitons in a monolayer semiconductor on twisted bilayer graphene

Minhao He[1‡], Jiaqi Cai[1‡], Huiyuan Zheng[2‡], Eric Seewald[3], T. Taniguchi[4], K. Watanabe[5], Jiaqiang Yan[6,7], Matthew Yankowitz[1,8], Abhay Pasupathy[3], Wang Yao[2,9†], and Xiaodong Xu[1,8†]

[1]Department of Physics, University of Washington, Seattle, Washington, 98195, USA

[2]Department of Physics, University of Hong Kong, Hong Kong, China

[3]Department of Physics, Columbia University, New York, NY, 10027, USA

[4]International Center for Materials Nanoarchitectonics,

National Institute for Materials Science,

1-1 Namiki, Tsukuba 305-0044, Japan

[5]Research Center for Functional Materials,

National Institute for Materials Science,

1-1 Namiki, Tsukuba 305-0044, Japan

[6]Materials Science and Technology Division,

Oak Ridge National Laboratory, Oak Ridge, Tennessee, 37831, USA

[7]Department of Materials Science and Engineering,

University of Tennessee, Knoxville, Tennessee, 37996, USA

[8]Department of Materials Science and Engineering,

University of Washington, Seattle, Washington, 98195, USA

[9]HKU-UCAS Joint Institute of Theoretical and Computational Physics at Hong Kong, China

[‡] These authors contributed equally to the work. and

[†] wangyao@hku.hk (W.Y.); xuxd@uw.edu (X.X.)

**S1. Exciton hybridization induced by moiré potential**

In this paragraph, we discuss the possible origins of exciton hybridization in the WSe$_2$/twisted bilayer graphene (TBG) system.

In center-of-mass (COM) coordinate, the wavefunction of exciton with COM kinematic momentum $\mathbf{Q}$ in the TMD monolayers can be effectively modeled by:

$$\Psi_{\mathbf{Q}}(\mathbf{r}_e, \mathbf{r}_h) = \sum_{\mathbf{q}} \psi(\mathbf{q}) |c_{\mathbf{q}+\gamma_e \mathbf{Q}} v_{\mathbf{q}-\gamma_h \mathbf{Q}}\rangle$$

where the Bloch electrons and holes are located around the Brillouin zone (BZ) corners. Their momentum $\mathbf{k}_e$ and $\mathbf{k}_h$ can be recombined into the relative motion (RM) momentum $\mathbf{q}$ and the center-of-mass (COM) momentum $\mathbf{Q}$. The $\gamma_e \equiv m_e/(m_e + m_h)$ and $\gamma_h \equiv m_h/(m_e + m_h)$ describe their effective mass. The effective Hamiltonian contains the kinetic energy and Coulomb interaction:

$$H_0 = \frac{\hbar \mathbf{k}_e^2}{2m_e} + \frac{\hbar \mathbf{k}_h^2}{2m_h} + V(\mathbf{r}_e, \mathbf{r}_h)$$

If the interaction between the electron and hole has the translational symmetry $V(\mathbf{r}_e, \mathbf{r}_h) = V(\mathbf{r}_e - \mathbf{r}_h)$, the RM of the exciton is decoupled from the COM and has the hydrogen-like form, known as the exciton Rydberg states.

In the WSe$_2$/twisted bilayer graphene (TBG) system, given the large lattice mismatch between WSe$_2$ and graphene, the WSe$_2$/graphene interface can not introduce long-wavelength moiré potential on the excitons. Nevertheless, the periodic dielectric screening effect from the charge distribution in TBG patterned by its own moiré superlattice can affect the WSe$_2$ bandgap and exciton binding energy, therefore manifests as a superlattice potential for the WSe$_2$ exciton.

If the variation length scale of the moiré potential is much larger than the radius of the excitonic Rydberg states, the RM is essentially still decoupled from COM. Only the COM is affected by the superlattice potential, which causes Umklapp scattering and formation of the exciton minibands [1, 2]. The slowly varying moiré potential can be approximated by the harmonic approximation, which means the Umklapp scattering is mainly contributed by the low-order moiré reciprocal lattice vectors. However, two experimental facts exclude the Umklampp scattering as the origin of the observed new resonance peaks with the increasing doping: i) The umklampp peak of the 1s state is absent experimentally. 2) The relative peak positions are independent of the twist angle, i.e., the variation length scale of moiré potential.

In fact, the charge distribution in TBG upon doping is highly localized as measured in STM experiments, so it provides a periodic dielectric environment that rapidly varies in space (Fig. **1b** in the main text). The dielectric environment can affect exciton Rydberg states differently, depending on the relative scale of their radius compared to the moiré potential variation. The 1s ground state has the smallest radius [3], and the observation of a nearly independent 1s resonance as doping varies suggests that the dielectric screening effect on the quasiparticle band gap and exciton binding energy still cancels, as observed in other context of dielectric variations [2]. For higher Rydberg states with significantly larger radius [3], the highly localized charge distribution produces a spatially steep dielectric function, such that the RM can no longer be decoupled from the COM, and the otherwise orthogonal Rydberg states can get hybridized by the deep moiré potential.

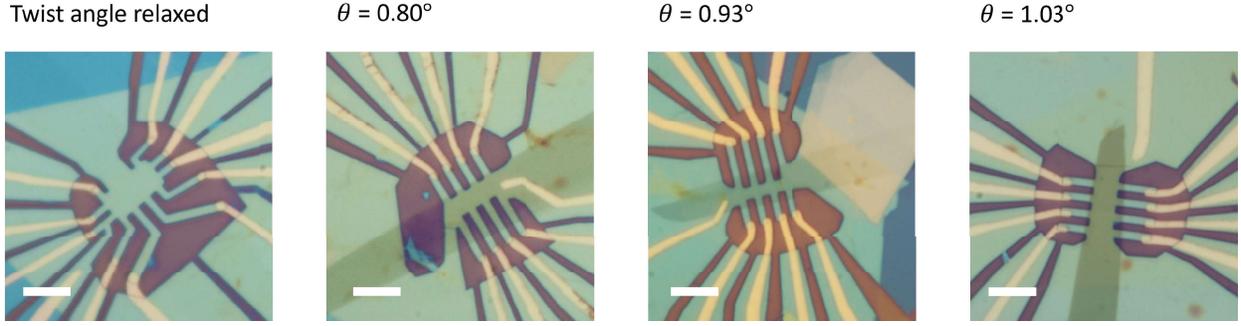

**Extended Data Figure 1. Optical images of four WSe$_2$/TBG devices in this study.** All scale bars represent 6 $\mu$m.

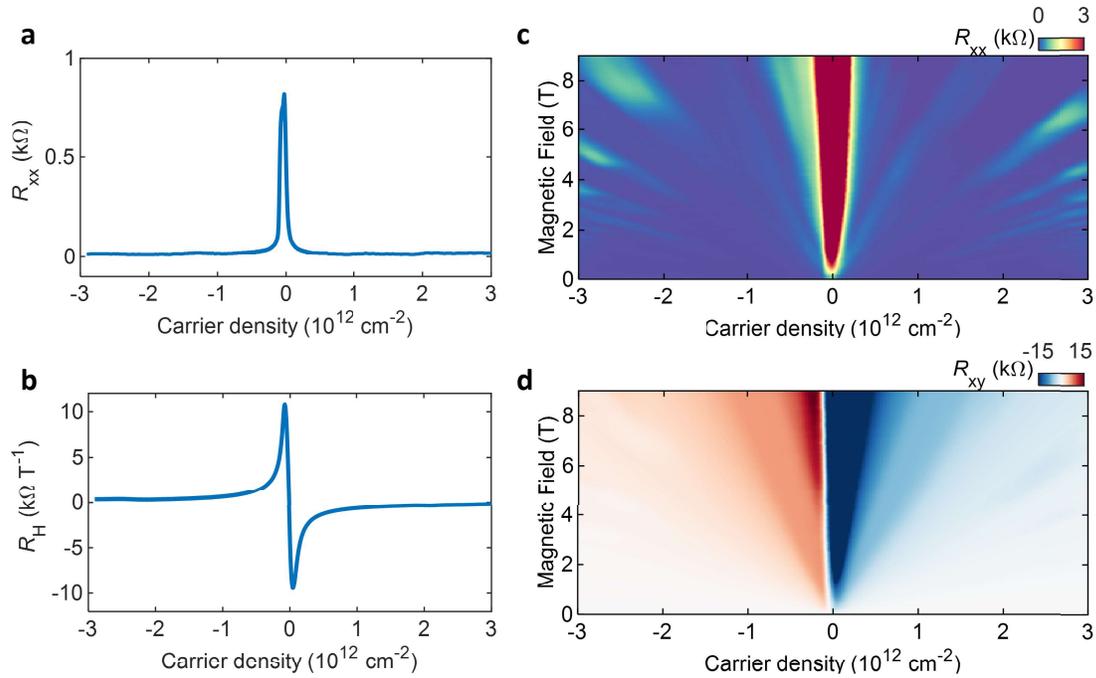

**Extended Data Figure 2**. **Transport properties of WSe$_2$/TBG with TBG twist angle relaxed.** **a-b,** Resistance ($R_{xx}$) and Hall coefficient ($R_H$) as a function of carrier density measured in a WSe$_2$/TBG device. **c-d,** Landau fan diagram of the resistance ($R_{xx}$) and Hall resistance ($R_{xy}$) respectively, measured in the same device. The absence of Hofstadter states and Brown-Zak Oscillation, together with absence of carrier type reversal shown in **b**, all point to a relaxed twist angle for the TBG. All measurements are performed at a temperature of $T = 2$ K.

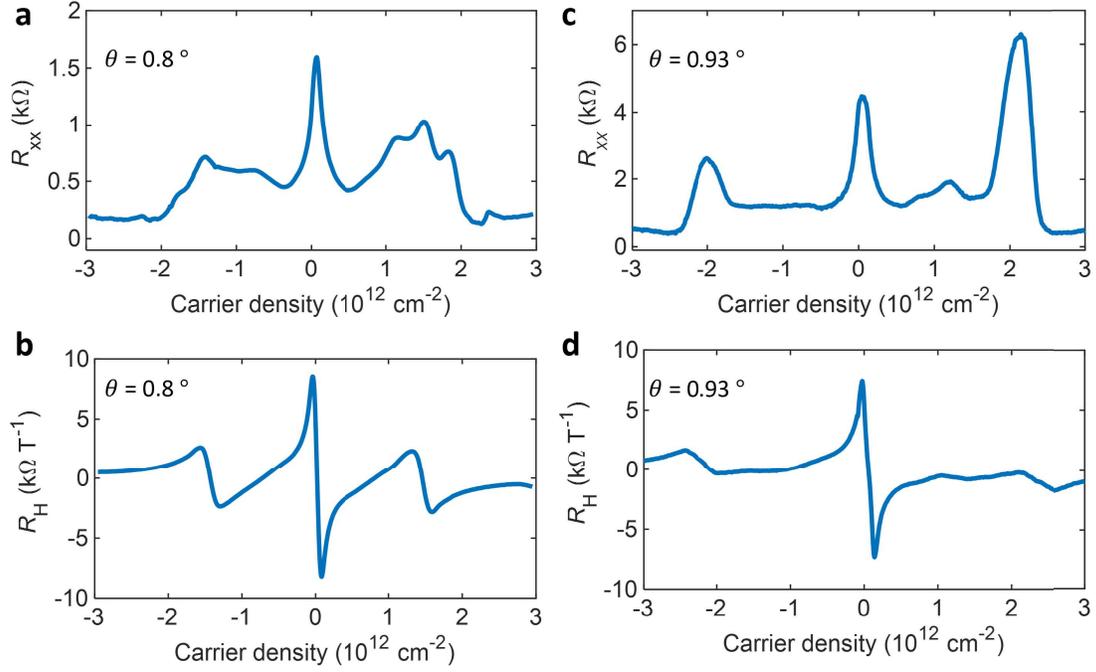

**Extended Data Figure 3. Transport properties of WSe$_2$/TBG with $\theta = 0.8°$ and $\theta = 0.93°$. a,c** Resistance ($R_{xx}$) and **b,d** Hall coefficient ($R_H$) measured in two WSe$_2$/TBG devices respectively. The concomitant resistive states in $R_{xx}$ and the carrier type reversals mark the band gaps between the lowest flat bands and the remote bands. Their corresponding twist angle are calculated to be $\theta = 0.8°$ and $\theta = 0.93°$ respectively. All measurements are performed at a temperature of $T = 2$ K.

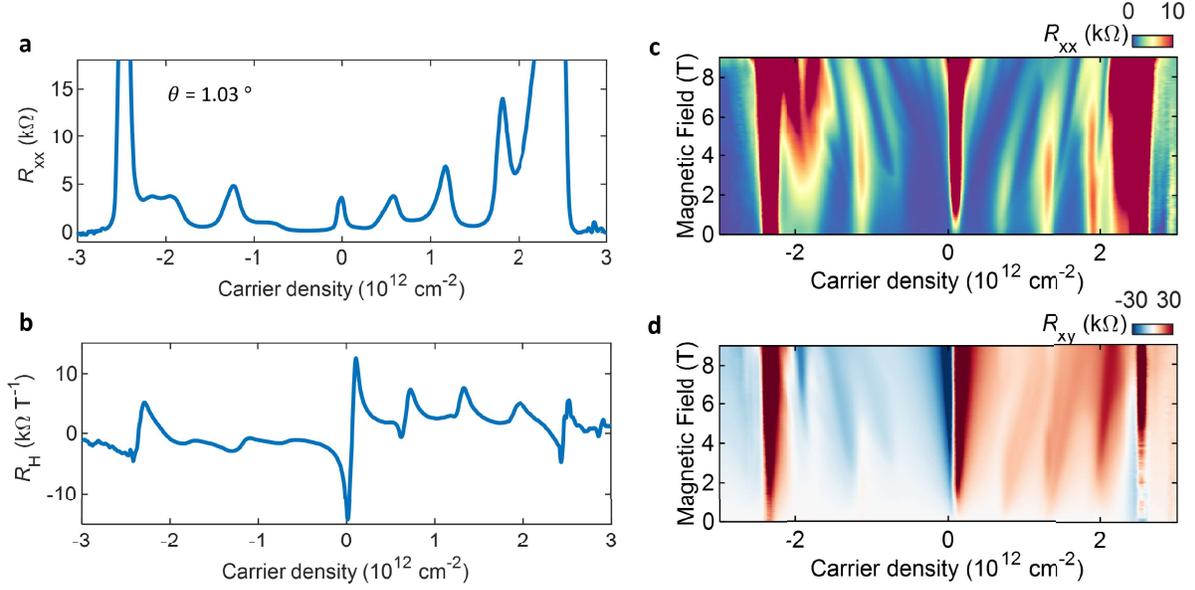

**Extended Data Figure 4**. Transport properties of WSe$_2$/MATBG with $\theta = 1.03°$. **a-b,** Resistance ($R_{xx}$) and Hall coefficient ($R_H$) as a function of carrier density measured in the WSe$_2$/MATBG device. We observed correlated states at integer filling factor $\nu$ = -3, -2, 0, 1, 2, 3 in addition to the band insulators at filling factor $\nu = \pm 4$. **c-d,** Landau fan diagram of the resistance ($R_{xx}$) and Hall resistance ($R_{xy}$) respectively, measured in the same device. We observe a typical cascade sequence of the high field correlated Chern insulators as frequently reported in MATBG [4–10]. The twist angle of the MATBG is determined to be $\theta = 1.03°$ as determined from both the Hall coefficient and the Hofstadter states. All measurements are performed at a temperature of $T = 2$ K.

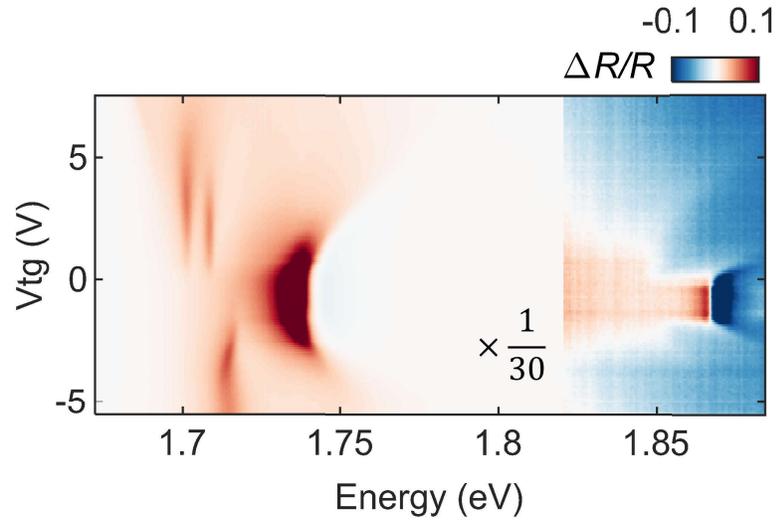

**Extended Data Figure 5**. **Reflectance contrast of a bare monolayer WSe$_2$** Reflectance contrast gate map ($\Delta R/R$) of a monolayer WSe$_2$ without adjacent graphene layers. We observe 1s exciton, 2s Rydberg exciton, and their corresponding charged excitons, consistent with literature [11]. This measurement shows the high quality of our device.

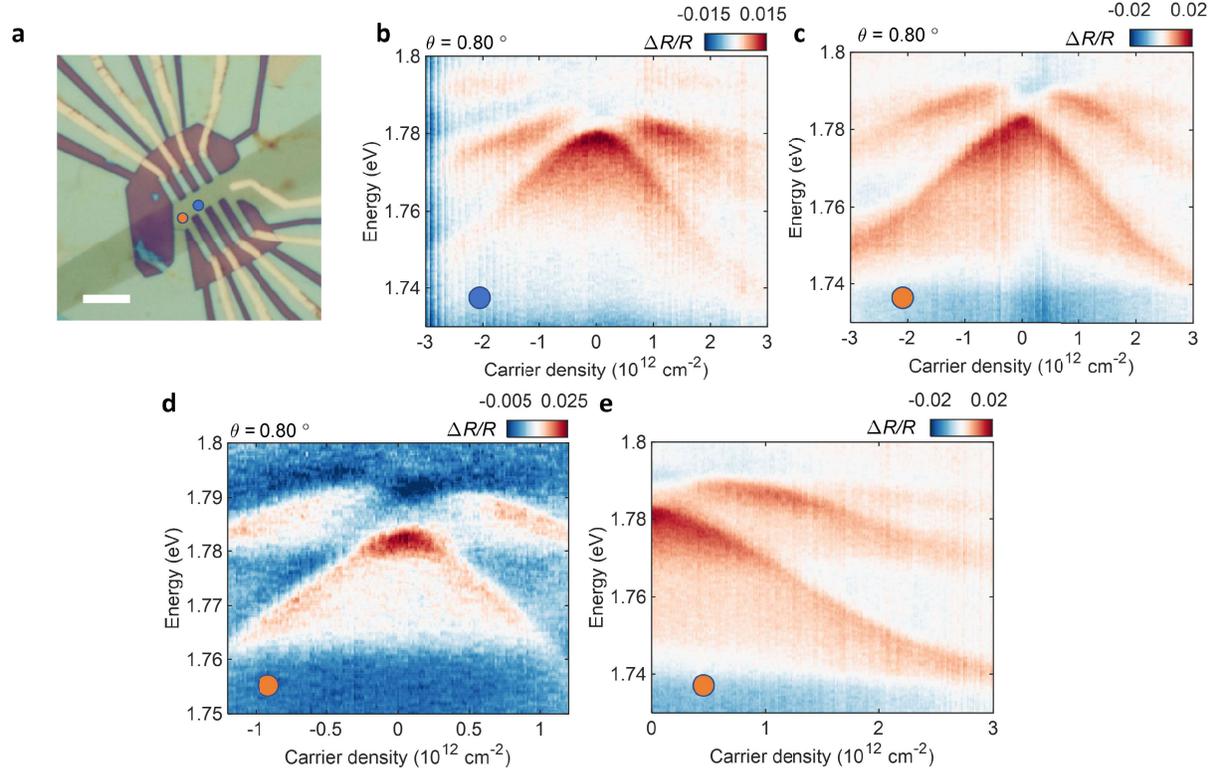

**Extended Data Figure 6**. Reflectance contrast in the WSe$_2$/TBG $\theta = 0.80°$ device. **a,** Optical image of the WSe$_2$/TBG $\theta = 0.80°$ device. Scale bar represent 6 $\mu$m. **b-c,** Reflectance contrast map $\Delta R/R$ measured at different spots in the same sample, labelled by blue and orange dots in **a** respectively. This measurement shows the reproducibility of the moiré excitons. **d,** High resolution reflectance contrast map $\Delta R/R$ with a focus on the first emergent moiré exciton. This measurement emphasises the discontinuous nature of the original 2s Rydberg state and its moiré induced hybridized exciton, with a color scale off-centered from zero. **e,** High resolution reflectance contrast map $\Delta R/R$ focusing on the electron doping side. This measurement emphasises the sequential emergence of the moiré excitons with increasing charge doping in TBG.

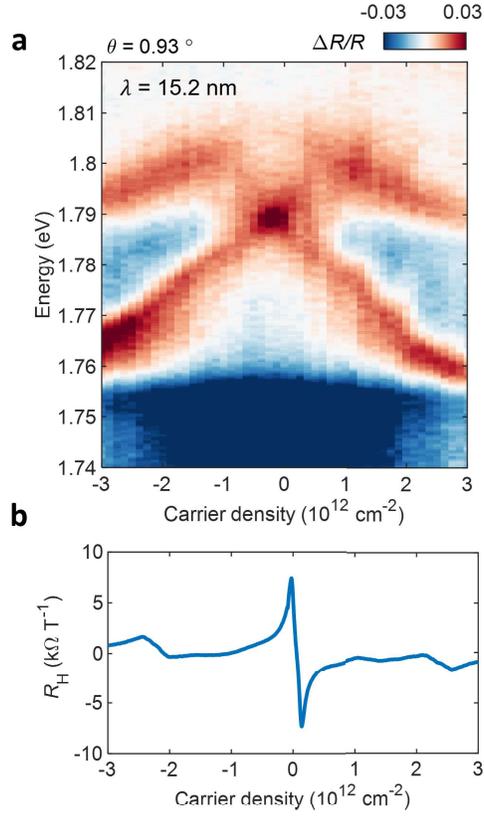

**Extended Data Figure 7. Reflectance contrast in the WSe$_2$/TBG $\theta = 0.93°$ device. a,** Reflectance contrast map $\Delta R/R$ of the WSe$_2$/TBG $\theta = 0.93°$ device. Similar to observation in the $\theta = 0.80°$ device, the higher energy moiré exciton emerges at finite charge doping of the TBG. **b,** Hall coefficient $R_H$ as a function of carrier density in the same $\theta = 0.93°$ device.

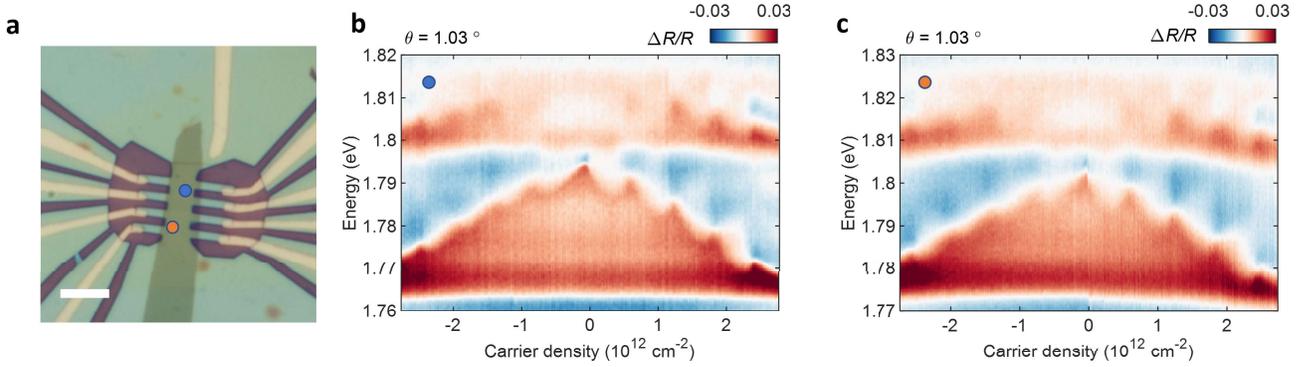

**Extended Data Figure 8. Reflectance contrast in the WSe$_2$/MATBG $\theta = 1.03°$ device. a,** Optical image of the WSe$_2$/MATBG $\theta = 1.03°$ device. Scale bar represent 6 $\mu$m. **b-c,** Reflectance contrast map $\Delta R/R$ measured at different spots in the same sample, labelled by blue and orange dots in **a** respectively. This measurement shows the high uniformity of the sample and the reproducibility of optical sensing of the correlated states in MATBG.

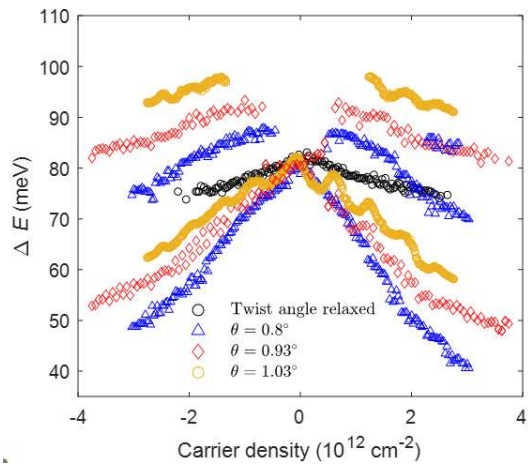

**Extended Data Figure 9. Twist angle dependence of the moiré Rydberg excitons' energy.** Summary of the 2s Rydberg exciton and the moiré Rydberg excitons' energy of all four devices measured. Their energy is plotted as a relative energy $\Delta$E as compared with their corresponding $E_{1s}$.

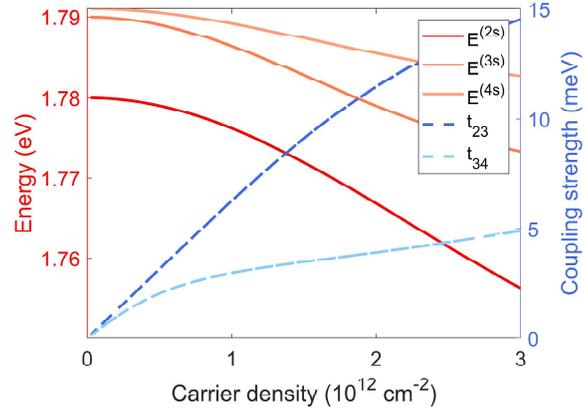

**Extended Data Figure 10**. **Rydberg states energy and coupling strength used in simulation.** Rydberg states energy $E^{Ns}(n)$ and coupling strength $t_{23}(n)$, $t_{34}(n)$ plotted as function of carrier density $n$ in TBG.